\documentclass[12pt]{article}
\usepackage{amssymb,graphicx}
\newcommand{\be}{\begin{equation}}
\newcommand{\ee}{\end{equation}}
\newcommand{\bd}{\begin{displaymath}}
\newcommand{\ed}{\end{displaymath}}
\newcommand{\ba}{\begin{eqnarray}}
\newcommand{\ea}{\end{eqnarray}}
\newcommand{\bp}{\left(}
\newcommand{\ep}{\right)}
\newcommand{\bb}{\left[}
\newcommand{\eb}{\right]}

\newcommand{\nn}{\nonumber\\}
\newcommand{\im}{\textrm{Im}}

\renewcommand{\k}{\textbf{k}}
\newcommand{\kabs}{|\textbf{k}|}
\newcommand{\xvon}{\textbf{x}}
\newcommand{\arctg}{\textrm{arctg}}
\def\textmini#1{\textrm{\small #1}}

\begin{document}
\title{Relaxation of $2+1$ dimensional
classical O(2) symmetric scalar fields}

\author{{Sz. Bors\'anyi\footnote{mazsx@cleopatra.elte.hu}~
 and Zs. Sz{\'e}p\footnote{szepzs@antonius.elte.hu}}\\
{Department of Atomic Physics}\\
{E{\"o}tv{\"o}s University, Budapest, Hungary}\\
}
\vfill
\maketitle
\begin{abstract}
Real time thermalization and relaxation phenomena are studied
in the low energy density phase of the 2+1 dimensional
classical O(2) symmetric scalar theory by solving numerically its dynamics. 
The near-equilibrium decay rate of on-shell waves and the power law governing
the large time asymptotics of the off-shell relaxation agree with the
analytic results based on linear response theory. The realisation of
the Mermin-Wagner theorem is also studied in the final equilibrium ensemble.
\end{abstract}


The approach to equilibrium of initially out-of-equilibrium states is
a highly important issue in many branches of physics ranging 
from inflationary cosmology
(the spectrum of density fluctuation in the early universe)
through particle physics (the problem of bariogenesis, formation of
DCC in heavy ion collisions) to statistical physics (dynamics of phase
transitions, realization of Boltzmann's conjecture \footnote{
An ensemble of isolated interacting systems approaches thermal
equilibrium at large times.}.

In the recent field theoretical literature thermalization and
relaxation of classical fields is intensively investigated 
 \cite{parisi98,aarts00}.
Its interest follows from the presence of bosonic degrees of freedom 
with high occupation numbers, for instance, in cosmological
applications \cite{felder00}.
In fact, efficient methods for the determination of the exact,
real time evolution, e.g. numerical integration of the
equations of motion are known only for classical fields.
Truncation and expansion schemes should be benchmarked against this
exact solution.

Recently, evolution of equal-time 1PI correlation functions derived from the
effective time dependent action \cite{wetterich99}
were confronted with the results of the exact time evolution.
By solving equations non-local in time, obtained from 2PI effective
action thermalization of quantum fields was demonstrated \cite{cox}.
These approximate methods can be formulated both for
classical and quantum cases. In the quantum case the analogue of the
classical ensemble averaging over the initial
conditions is the quantum expectation
value. With this correspondence the formal derivations, and
hence the results, are quite similar \cite{aarts98}.

In statistical physics real time studies of complex nonlinear
systems have shown that
expectation values of observables are correctly reproduced by
averaging over the microcanonical evolution of a single equilibrium system,
even if a very small part of the phase space is covered by the motion
during the interval of observation \cite{Caiani98}.

\begin{figure}
\begin{center}
\includegraphics[width=8cm]{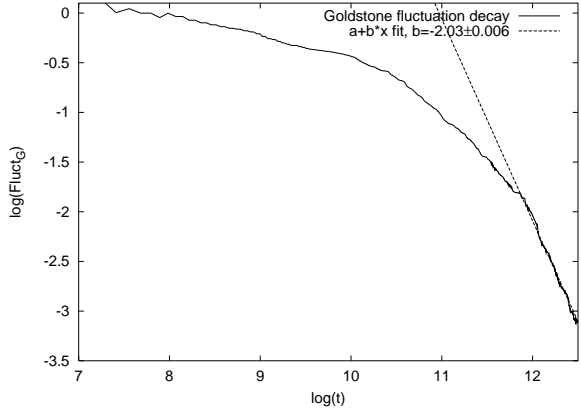}
\end{center}
\caption{Relaxation of Goldstone spatial fluctuations with time
on ($e$-based) log-log scale plot.
($256\times 256$ lattice, $h=0.0001/\sqrt{6}$, $m_G=(62\pm5)\cdot10^{-4}$,
$m_H=1.329\pm0.001\cdot10^{-3}$, the average over 69 runs is shown.)}
\label{goldstonedecay}
\end{figure}

In this Letter we wish to gain insight into the validity of the linear
response theory, by comparing its results to exact solutions. A similar
attempt was already done in \cite{aarts00}. There
the early, far from equilibrium time evolution was confronted with the
linear response results, and a relaxation slower than expected has
been observed. 
We will show later (see~Fig.~\ref{goldstonedecay}) that
the decay in the linear regime realizes the fastest relaxation
rate of this quantity during the whole time evolution. 
Exploration of the range of validity of the linear regime
is possible only by following the evolution very long,
and averaging over initial conditions in order to raise the signal
over the noise level. 

\textit{The model and its numerical solution.}
We concentrate on the $O(2)$ symmetric classical scalar theory with
a small explicit breaking term in its Lagrangian:
\be
{\cal L}=\frac12(\partial_\mu\Phi_1)^2+
\frac12(\partial_\mu\Phi_2)^2
-\frac12m^2\Phi_1^2-\frac12m^2\Phi_2^2
-\frac{\lambda}{24}\bp\Phi_1^2+\Phi_2^2\ep^2
+h\Phi_1,
\label{lagrange}
\ee
where $h$ controls the explicit symmetry breaking, and $m^2<0$. We have 
studied the time evolution of a system, discretized on lattice 
in two space dimensions at so low energy density (see below) that according
to the mean field analysis this would correspond to the broken symmetry 
phase. 

{The dynamics eventually drives the system towards thermal 
equilibrium. For 
$h\neq 0$, the equilibrium state has large magnetisation, nearly corresponding 
to the minimum of the classical potential. Fluctuations around this state
are naturally divided into Goldstone (light) and Higgs (heavy) excitations.
These excitations experience an effective potential, which agrees very well
with the result of finite volume perturbation theory. The relaxation into 
this state is the main subject of this paper. We shall present a detailed
comparison of the exact time evolution with the linear response theory.

A second relaxation process can be initiated from the equilibrium with
$h\neq 0$, if the magnetic field is switched off. The final $h=0$ equilibrium,
however, obeys in the thermodynamical limit the Mermin-Wagner-Hohenberg
theorem \cite{mermin}, which states the absence of spontaneous magnetisation
in two dimensional systems with continuous symmetry.  For equilibrium two
dimensional {\it finite volume} systems even for vanishing external source,
at very low temperature there is a non-zero magnetisation with a well
defined direction as shown in \cite{archambault}.
The finite volume magnetisation selects the direction with
respect to which we can define a parallel and a perpendicular mode. This
natural choice of base suggests the use of the Goldstone boson terminology
in the $h=0$ case too.
The way the finite magnetisation disappears as the volume goes to infinity
will be touched upon shortly at the end of our discussion.
}

The exact time evolution of the lattice system was studied by introducing
dimensionless field variables $\Phi_i\rightarrow\sqrt{a}\Phi_i$
and rescaling further these fields like
$\Phi^{\textmini{rescaled}}_{1,2}=\Phi_{1,2}\sqrt{\lambda/6}$
and $h^{\textmini{rescaled}}=h\sqrt{\lambda/6}$.
This sets the bare coupling to $\lambda^{\textmini{rescaled}}=6$.
We have chosen $m^2a^2=-1$, therefore all dimensionful quantities are 
actually expressed in units of $|m|$. Squared lattices of size
ranging from $64\times64$ to $512\times512$ were used. 

The explicit symmetry breaking parameter $h$ is chosen in the range
$0\dots0.0025/\sqrt{6}$. 
{Initially the system is at rest near the origin
of the $\Phi$-space. }
A small amplitude white noise determines the field in
every lattice point ($\Phi_1(\xvon,t=0)=\eta_1(\xvon),
\Phi_2(\xvon,t=0)=\eta_2(\xvon),
\dot\Phi_1(\xvon,t=0)\equiv\dot\Phi_2(\xvon,t=0)\equiv0,
\langle\eta_i(\xvon)\eta_j({\bf y})\rangle\sim\delta_{{\bf x},{\bf y}}
\delta_{i,j}$).
This means, that the space average of $\Phi_i$, the two-component order 
parameter (OP)
$(\overline{\Phi_1}^V,\overline{\Phi_2}^V)$ is very close initially to zero, 
which is
(approximately) a local maximum of the effective potential.
{The roll-down towards the minimum is oriented by the external
field $h$.} 
It is the potential energy difference of the initial and the final states,
which is redistributed between the modes.
The magnitude of the noise was so small, that the final energy density of the
system is exclusively determined by this difference in the potential energy.

{
Numerical diagonalization of the measured $2\times2$ matrix of equal
time fluctuations 
($C_{i,j}(t)=\overline{\Phi_i(\xvon ,t)\Phi_j(\xvon ,t)}^V-
\overline{\Phi_i(\xvon ,t)}^V
\overline{\Phi_j(\xvon ,t)}^V$) gave as uncorrelated degrees of freedom the
radial (Higgs) and angular (Goldstone) components of $(\Phi_1,\Phi_2)$:
\be
\Phi_H(\xvon,t)=\sqrt{\Phi_1^2(\xvon,t)+\Phi_2^2(\xvon,t)},\qquad
\Phi_G(\xvon,t)=\Phi_{H,0}\arctan\frac{\Phi_2(\xvon,t)}{\Phi_1(\xvon,t)},
\label{eq:felbontas}
\ee
$\Phi_{H,0}=|m|\sqrt{6/\lambda}+h/2|m|^2+{\cal O}(h^2)$ being the
minimum of the bare potential valley. This statement means
that $\overline{\Phi_H(t)\Phi_G(t)}^V\approx
\overline{\Phi_H(t)}^V\overline{\Phi_G(t)}^V$ for all $t$. 
The boundedness of $\Phi_2$ fluctuations follows
from the IR-regulating effect of the explicit symmetry
breaking term.
The position of the tree level minimum scaled by $|m|$ is 
$\Phi_{H,0}=1+h/2+{\cal
O}(h^2)$, and the scaled squared masses read as 
$m_H^2=2+h/2+{\cal O}(h^2)$ for the radial mode and $m_G^2=h+{\cal O}(h^2)$ 
for the angular one.
}

\begin{figure}
\begin{center}
\includegraphics[width=6cm]{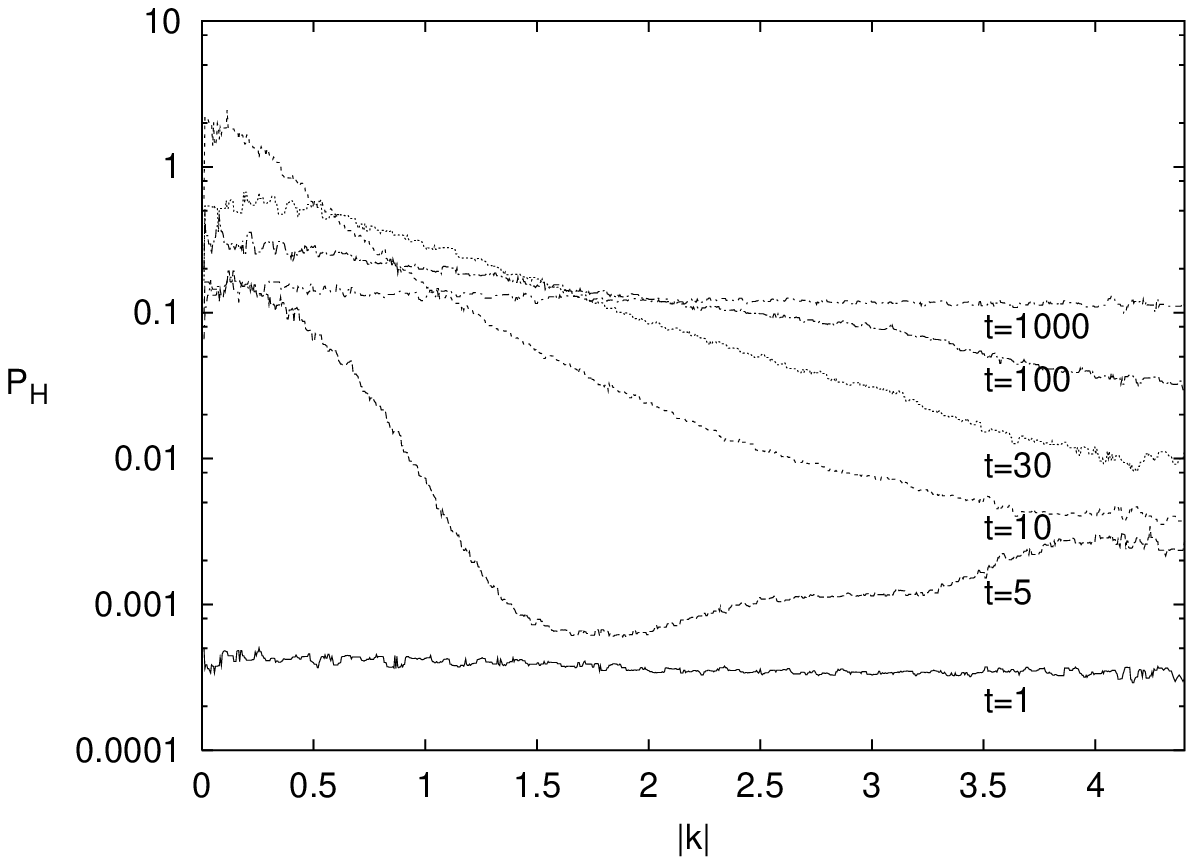}
\includegraphics[width=6cm]{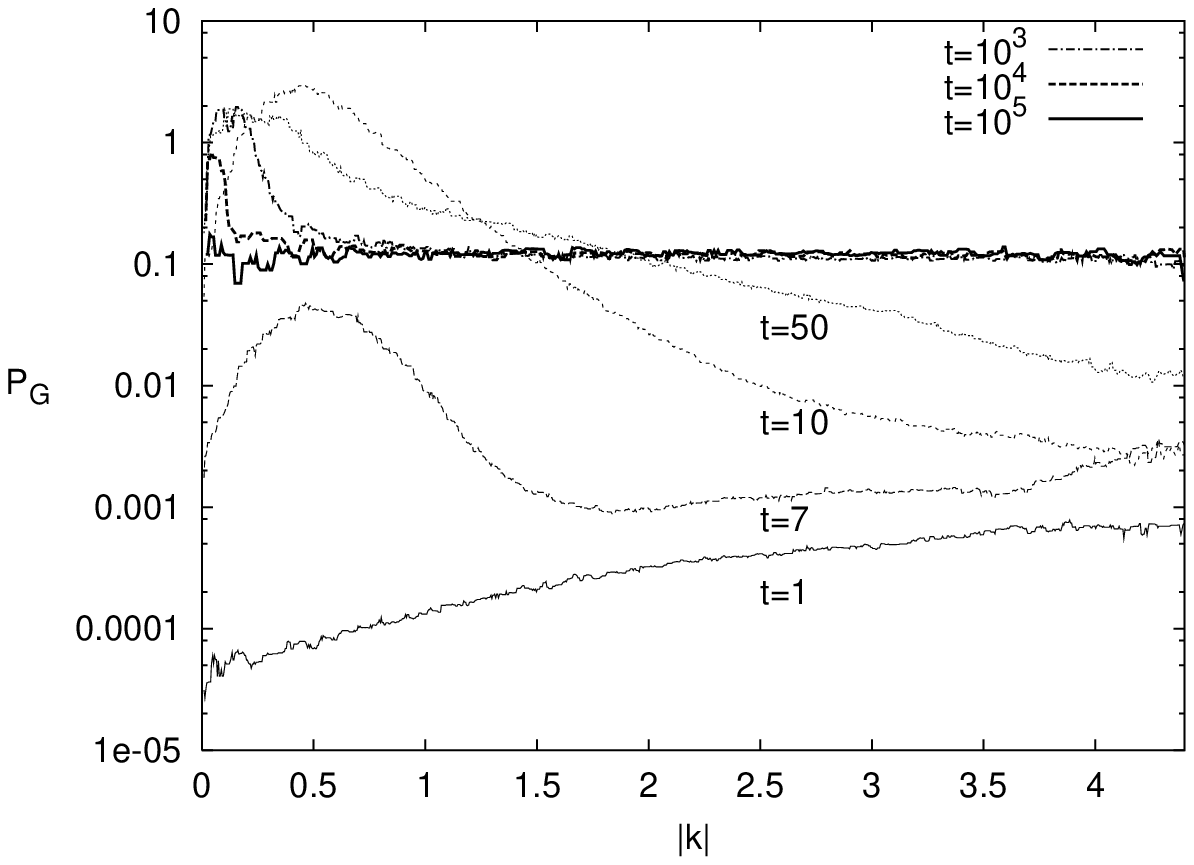}
\end{center}
\caption{Time evolution of $\kabs$ power spectra of Higgs (left) and Goldstone
(right) field components ($512\times512$ lattice,\, $h=0.0025/\sqrt{6}$ )}
\label{powerspectra}
\end{figure}

\textit{Relaxation to the $h\not=0$ equilibrium.}
During and immediately after the roll-down 
the main mechanism of the fast excitation of spatial
fluctuations is the parametric resonance, as observed and studied by
many authors, e.g. \cite{khlebnikov,greene97}. Also the dephasing of
different oscillation modes can be observed as reported in
\cite{wetterich99,mottola97}. The time scales of this stage are
characterised by the cutoff and by the masses of the radial and angular
modes. The measurement of these latter is discussed later.

The early time evolution can be characterised by the variation of the
kinetic power
spectra of the Higgs and Goldstone fields (see Fig. \ref{powerspectra}).
After dephasing the potential and kinetic energy of each
mode is balanced in itself, therefore the energy
content of different modes may be characterised by
{
the kinetic power spectrum
\be
P_{H,G}(\kabs)=\overline{{\dot\Phi_{H,G}}^2}^{\Omega_{\bf k}},
\ee
}where the averaging should be taken over the polar angle in momentum
space. 
In the final equilibrium state 
this quantity does not depend neither on $\kabs$,
{nor on which field it refers to}. Its value
corresponds to the temperature, which
{was measured} in our case to be
$P_{H,G}(\kabs)\equiv T^{kin}=0.125\pm0.0005$. 
This kinetic temperature can obviously be defined also
out-of-equilibrium
{as $T^{kin}=\overline{{\dot\Phi_{H,G}}^2}^{{\Omega_{\bf k}}}$}.

The initial parametric peak in the power spectrum of the Higgs field 
disappears rapidly. It proceeds mainly through decays into
pairs of the lighter Goldstone modes. This damping is reflected by an
energy transfer from the Higgs to the Goldstone modes.
In addition, the excitation of spinodally unstable Higgs modes also
contributes to the relaxation, as it was shown in \cite{maxwell}.

The parametric low $\kabs$ Goldstone peak, however, survives as long as
$t=10^4\dots 10^5$. During its long relaxation the energy is slowly
transferred back to the Higgs modes.  This process qualitatively corresponds
to off-shell emission of Higgs waves. The final equilibrium is reached
in a slow, non-exponential relaxation of $T^{kin}_G$ to $T^{kin}_H$.

The mechanism sketched above is suggested by the structure of the
analytic formulae derived in the perturbative relaxation
analysis, which should be relevant at least to the large time asymptotics
of the evolution. The forward time evolution of an initial
configuration is determined  by the classical self-energy function
$\Pi (\xvon,t)$:
\be
\Phi_{H,G}(t,\k)=\int\limits^{\infty}_{-\infty}\frac{dk_0}{2\pi}
\frac{z(\k)}{k^2-m_{H,G}^2-\Pi_{H,G}(k)}e^{-ik_0t},
\label{forward}
\ee
where $z(k)=ik_0F(\k)-P(\k)$, determined by the corresponding
initial configuration
\hbox{$F(\k)=\Phi(t=0,\k)$}, \hbox{$P(\k)=\partial_t\Phi(t=0,\k)$}.

For the calculation of the Fourier transform of $\Pi (\xvon ,t)$ 
the procedure described in \cite{aarts97,jako97} was used. 
(For the quantum treatments, which lead for
large occupation numbers to the same results, see
\cite{boyanovsky,mrowczinski}). 

First one makes the shift
$\Phi_1\to\Phi_1+\bar\Phi$ in order to describe the constant equilibrium
background, where $\bar\Phi$ is the classical ensemble average of
$\Phi_H(\xvon)$. Its direction is actually selected by the explicit 
symmetry breaking. Then in the linear approximation one finds
$\Phi_H(\xvon)\approx\Phi_1(\xvon)$ 
{and 
$\Phi_G(\xvon)\approx\Phi_2(\xvon)\Phi_{H,0}/{\bar\Phi}$.
We refer to our previous publication \cite[Appendix B]{jako99}
}
for the detailed derivation of the self energies, which read now for the Higgs
and Goldstone fields, as follows:
\ba
\Pi_{H}(k)
&=&-\bp\frac{\lambda\bar\Phi}{3}\ep^2T\int\frac{d^2q}{(2\pi)^2}
\int\frac{d^2p}{(2\pi)^2}\int d^2xdte^{i(k_0 t-{\bf k}{\bf x})}
e^{i({\bf q}-{\bf p}){\bf x}}\Theta(t)\times\nonumber\\
&&\times\bb
\frac{9\sin \omega_{H,{\bf q}}t}{\omega_{H,{\bf q}}}
\frac{\cos \omega_{H,{\bf p}}t}{\omega_{H,{\bf p}}^2}+
\frac{\sin \omega_{G,{\bf q}}t}{\omega_{G,{\bf q}}}
\frac{\cos \omega_{G,{\bf p}}t}{\omega_{G,{\bf p}}^2}
\eb,\\
\Pi_{G}(k)
&=&-\bp\frac{\lambda\bar\Phi}{3}\ep^2T\int\frac{d^2q}{(2\pi)^2}
\int\frac{d^2p}{(2\pi)^2}\int d^2xdte^{i(k_0 t-{\bf k}{\bf x})}
e^{i({\bf q}-{\bf p}){\bf x}}\Theta(t)\times\nonumber\\
&&\times\bb
\frac{\sin \omega_{G,{\bf q}}t}{\omega_{G,{\bf q}}}
\frac{\cos \omega_{H,{\bf p}}t}{\omega_{H,{\bf p}}^2}+
\frac{\sin \omega_{H,{\bf q}}t}{\omega_{H,{\bf q}}}
\frac{\cos \omega_{G,{\bf p}}t}{\omega_{G,{\bf p}}^2}
\eb,
\ea
where $\omega_{H/G,{\bf k}}^2=m_{H/G}^2+|{\bf k}|^2$. 
{The relevant mass values were calculated numerically from the
relaxed field configurations with the method developed in \cite{maxwell}.}
At the parameter values specified in our simulations the typical masses
are $a m_H\approx 1.3$ and $a m_G\lesssim 0.033$.

The imaginary part of the self energy, which accounts for the damping
phenomena, may be extracted using the principal value theorem.
There is a technically important difference between our present formulae 
and the ones derived in \cite{bors99}.
It follows from the fact
that our system is defined in two space dimensions. This circumstance
modifies the volume element in the momentum space and an
integrand of more complicated analytical structure appears
{
when calculating the imaginary part of the classical self energy function.
}

The explicit expressions of the imaginary parts are the following:
\ba
\im\Pi_H
&=&-\bp\frac{\lambda\bar\Phi}{2}\ep^2\Theta(-k^2)\frac{2T}{\pi}
\frac{1}{R_H}
\bp\frac{\pi}{2}
-\arctg\frac{R_H}{|k_0|\sqrt{-k^2}}\ep\nn
&&-\bp\frac{\lambda\bar\Phi}{6}\ep^2\Theta(-k^2)\frac{2T}{\pi}
\frac{1}{R_G}
\bp\frac{\pi}{2}
-\arctg\frac{R_G}{|k_0|\sqrt{-k^2}}\ep\nn
&&-\bp\frac{\lambda\bar\Phi}{2}\ep^2\Theta(k^2-4m_H^2)
\frac{T}{R_H}-\bp\frac{\lambda\bar\Phi}{6}\ep^2\Theta(k^2-4m_G^2)
\frac{T}{R_G},
\label{kiintegraltpihiggs}\\
\im\Pi_G&=&
-\bp\frac{\lambda\bar\Phi}{6}\ep^2 T
\bp P_G+P_H\ep\Theta(k^2)\Theta(k^2-(m_H+m_G)^2)\nn
&&-\bp\frac{\lambda\bar\Phi}{6}\ep^2 T
\bp P_G-P_H\ep
\Theta(k^2)\Theta((m_H-m_G)^2-k^2)\nn
&&-\bp\frac{\lambda\bar\Phi}{6}\ep^2
\frac{2T}{\pi} \Theta(-k^2)\left[ P_H
\arcsin\frac{k_0(m_G^2-m_H^2-k^2)}{\kabs\sqrt{\Delta}}\right.\nn
&&\left.\qquad\qquad+P_G
\arcsin\frac{k_0(m_H^2-m_G^2-k^2)}{\kabs\sqrt{\Delta}}\right],
\label{kiintegraltpigoldstone}
\ea
with
\ba
&\Delta=(k^2-m_H^2+m_G^2)^2-4m_G^2k^2,&\nn
&R_{H,G}=\sqrt{k^4+4\kabs^2m_{H,G}^2},\qquad
P_{H,G}=1/\sqrt{\Delta+4m_{H,G}^2k_0^2}.&
\ea

It is the off-shell damping coming from the contributions of the
cuts to (\ref{forward}) which has a direct impact on the
{late} 
time evolution of the OP we are analysing. We followed the method, developed in
\cite{boyanovskyb,boyanovsky}, which determines the leading power law
tail of the relaxation.
{
Contrary to the $3+1$ dimensional case in $2+1$ dimensions
the integrand of Eq. (\ref{forward}) has extra cuts. These can,
however, be directed in a way that either they do not contribute
(Goldstone case) or their contribution is suppressed by a factor of
$e^{-m_Ht}$ (Higgs case). The large time asymptotics of the two field
components is the following:
}
\ba
\Phi_H(t,\k)&=&
\frac{T}{\pi t}
\bp\frac{\lambda\bar\Phi}{6}\ep^2
\left[
\frac{1}{m_H^5}
\bp
\frac{P(\k)\cos(t\Omega_{H,{\bf k}})}{\Omega_{H,{\bf k}}}
-F(\k)\sin(t\Omega_{H,{\bf k}})
\ep\right.\nn
&&+\left.
\frac{1}{m_G(4m_G^2-m_H^2)^2}
\bp
\frac{P(\k)\cos(t\Omega_{G,{\bf k}})}{\Omega_{G,{\bf k}}}
-F(\k) \sin(t\Omega_{G,{\bf k}})
\ep\right]\nn
&&-
\frac{T}{(2\pi t)^{3/2}}
\bp\frac{\lambda\bar\Phi}{6}\ep^2
\frac{2\sqrt{\kabs}}{m_H^4}
\bp\frac{9}{m_H^2}+\frac{1}{m_G^2}\ep\times\nn
&&\quad\times\bp \frac{P(\k)}{\kabs}\cos(\kabs t-\frac{\pi}{4})
-F\sin(\kabs t- \frac{\pi}{4})\ep,\\
{\bar\Phi\over {\Phi_{H,0}}}\Phi_G(t,\k)&=&
-\bp\frac{\lambda\bar\Phi}{6}
\ep^2
\frac{T}{\pi t}
\left[
\frac{1}{m_H^2(m_H-2m_G)^2}
\bp\frac{1}{m_G}-\frac{1}{m_H}\ep\right.\nn
&&\times
\bp
\frac{P(\k)\cos \Omega_{H-G,\k}t}{\Omega_{H-G,\k}}
- F(\k) \sin t\Omega_{H-G,\k}
\ep\nn
&&+\frac{1}{m_H^2(m_H+2m_G)^2}
\bp
\frac{1}{m_G}+\frac{1}{m_H}
\ep\nn
&&\left.\times\bp
\frac{P(\k)\cos \Omega_{H+G,\k}t}{\Omega_{H+G,\k}}
- F(\k) \sin t\Omega_{H+G,\k}
\ep\right],
\label{goldtimedep}
\ea
where the threshold frequency values, appearing above are given by
$\Omega_{H/G,{\bf k}}^2={\bf k}^2+m^2_{H/G}$ and
$\Omega_{H\pm G,{\bf k}}^2={\bf k}^2+(m_H\pm m_G)^2$.

In order to obtain numerical evidence for the presence of the predicted
power law tail, we measured the quadratic spatial fluctuation moments
of both fields as a function of time. They are defined as
$Fluct(t)=\overline{\Phi(\xvon,t)^2}^V-\bp\overline{\Phi(\xvon,t)}^V\ep^2$.
Substituting (\ref{goldtimedep}) into this definition we get $\sim
t^{-2}$ like relaxation for late times. This 
behaviour is indeed observed as the fitted {power} shows
in Fig. \ref{goldstonedecay} for the example of the Goldstone mode.

The formulation of the theory upon which the above formulae were obtained
actually uses ensemble averages over the initial data of the classical
evolution. We found,
that although every single run seems to relax even quantitatively in the
same way, this late time evolution can be extracted from the noisy
data only, if this averaging 
is performed indeed (116 runs were involved).
We could
recognise a $\sim1/t^2$ decay for Higgs modes too, but because of the
low signal/noise rate we could not do quantitative analysis.
In particular the analysis is made difficult by the fact, that
Higgs modes decay rapidly and vanish in the noise before the power law
tail would be reached.

Relaxation behaviour of OP ($\Phi_H(${\bf k}=0$)$) was also investigated, and
--- in accordance with the expectation above --- an oscillation damped
by $\sim t^{-1\pm0.02}$ was found. Moreover, this oscillation, is observed
around a value, monotonically approaching its equilibrium value as
$\sim t^{-2\pm0.01}$. This latter behaviour is explained by the fact,
that the exact equation of motion for OP contains $Fluct(t)$ as a time
dependent parameter \cite{khlebnikov,maxwell}.
Its $\sim t^{-2}$ like behaviour is inherited by
the slowly varying part of OP. (The errors of the exponents come from
averaging over 92 runs on a $256\times256$ lattice.)

\textit{The $h\not=0$ equilibrium.}
When one follows the evolution of the system for long times
($t=10^6$), there seems no further relaxation to take place, i.e. the initial
signal has been lost in the thermal noise. In order to convince ourselves of
having arrived to thermal equilibrium we compare the measured values
of $\overline{\Phi_H}^V$, and the masses  $m_G$ and $m_H$ to the analytical
estimates of these quantities coming from the one-loop lattice effective
potential evaluated for the same size lattice as used in the
simulation. In the analytic expression we used the measured kinetic
temperature. The equilibrium masses $m_G$ and $m_H$ were determined in
the simulation both from the corresponding correlation functions and
by fitting the oscillatory motion around the equilibrium of the
corresponding OP-components as described in \cite{maxwell}.

We have checked that the system with the present initial conditions
is deeply in the Coulomb phase
i.e. the temperature was about four times smaller than the
Kosterlitz-Thouless critical temperature $T_{KT}$. In the absence of explicit
symmetry breaking on the critical line between $T=0$ and $T_{KT}$ the
Goldstone correlation length is expected to diverge with the lattice size.
The explicit symmetry breaking parameter $h$ acts as an
infrared regulator. Indeed, the measured Goldstone
correlation length is found to be proportional to $L$, but for 
the largest sizes ($L=256$), where the IR cutoff $h$ begins to dominate.
On the other hand, the inverse correlation length in the Higgs
direction is finite, its value being $3\%$ smaller than the two-loop
mass and $6\%$ smaller than the one-loop value.

A very good agreement of $\overline{\Phi_H}^V$ with
$\bar\Phi$ is found where $\bar\Phi$ is given by the minimum of the 
effective potential in the radial direction, the relative deviation being 
${\cal O}(10^{-4})$. 
The discrepancy between the measured (with the method described in
\cite{maxwell})
and calculated masses was less than $1\%$ 
and $5\%$ for the Higgs and Goldstone modes, respectively, the perturbative 
values being systematically smaller. Our measurements of the Goldstone mass 
became very noisy for small values of $h\lesssim0.00001/\sqrt{6}$. 

In the equilibrium system one can proceed to ``experiments'', which
check the correctness of the on-shell decay rates computed in the
linear response theory. These are the simple zeros of the denominator
of Eq.(\ref{forward}) which determine the exponential
damping of on-shell excitations. 
The damping rates are obtained by substituting $\kabs=k_0$ into
Eqs. (\ref{kiintegraltpihiggs}) and (\ref{kiintegraltpigoldstone}).
To leading order in $(\lambda\bar\Phi/6)^2T$ the rates read
(assuming $m_H>2m_G$) as
\ba
&&\Gamma_H=-\frac{\im\Pi_H(k^2=m_H^2)}{2\omega_{H,\bf k}}=
\bp\frac{\lambda\bar\Phi}{6}\ep^2\frac{T}{2\omega_{H,\bf k}}
\frac{1}{\sqrt{m_H^4+4\kabs^2m_G^2}}\\
&&\Gamma_G=\bp\frac{\lambda\bar\Phi}{6}\ep^2
\frac{T}{2\omega_{G,{\bf k}}}
\left[
\frac{1}{\sqrt{(m_H^2-2m_G^2)^2+4m_G^2\kabs^2}}
-\frac{1}{m_H\sqrt{m_H^2+4\kabs^2}}
\right].
\ea

\begin{figure}
\begin{center}
\includegraphics[width=8cm]{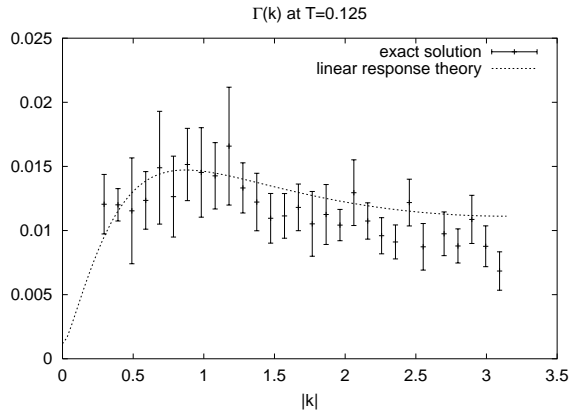}
\end{center}
\caption{Numerical and analytical values for Goldstone on-shell decay
rate as a function of $\kabs$
($128\times128$ lattice, $h=0.0025/\sqrt{6}$, $m_G=0.0319$,
$m_H=1.33$, $\bar\Phi=0.96$, $T=0.125$)}
\label{gammacomparison}
\end{figure}

{On equilibrium configurations single $\bf k$-modes have been 
superimposed with an amplitude between $0.01\dots0.2$.}
Time evolution of these excited modes showed perfect
exponential decay. The exponents did, however, depend on the
amplitude. For small amplitudes the fit was unstable, the mode was quickly
lost in the noise. For bigger values  nonlinear effects did show
up. Therefore we were looking for a plateau in 
the amplitude dependence of the decay rates between these two extremes.
Then its value has been compared to the analytical estimates (see
Fig.\ref{gammacomparison}). 
The error bars show how well-defined plateau could be found.
This numerical experiment was performed both for Higgs
and Goldstone waves, but the errors for Higgs decay were too large
again. The value of $\Gamma_H$ computed from the linear response theory
was reproduced within an accuracy of
$\pm50\%$, but we can not say anything about its predicted $\k$-dependence.
The reason for this seems to be, that the evolution becomes very
soon nonlinear as we try to increase the excitation amplitude
higher that the noise level.

{
As was indicated in \cite{hasenfratz} finite size effects may be highly
important in systems with Goldstone-like excitations. This
circumstance made necessary the use of finite volume perturbation
theory in the analysis above.
Neither the lattice summed perturbative effective potential, nor
the numerically found mass values showed any $L$ dependence if $L>64$.
We have checked for the $L$ independence in
all cases where the exact evolution and linear analysis were
confronted (e.g. the values of damping rates).

\textit{Diffusion to the $h=0$ equilibrium.}
The most care must be taken when the macroscopic
magnetisation $M(L)$ is considered. 
For finite volume,
the ensemble average computed from the low temperature spin-wave 
approximation is non-zero \cite{archambault} even in the absence of explicit 
symmetry breaking:
\be
M^2(L)\equiv\left(\overline{\Phi_1}^V\right)^2+
\left(\overline{\Phi_2}^V\right)^2 
\approx (2L)^{-T/(2\pi\bar\Phi^2)}.
\ee
For each
realization of the canonical ensemble observed in Monte Carlo simulations 
the magnetisation vector has a well-defined direction
$\theta =\arctan\left(\overline{\Phi_2}^V/\overline{\Phi_1}^V\right)$.

It is an interesting question, by what mechanism the MWH-theorem is realized 
as $L\to\infty$. 
It was shown in Monte Carlo simulations that the erasure of the order 
is realized by the diffusion like displacement of
the direction of $\theta$ \cite{archambault}.

In our real time study, 
for the investigation of the onset of the finite volume version of
MWH-theorem a number of runs were continued from the equilibrium state
reached with $h\neq 0$, after
the magnetic field was  
switched off. In each individual system OP begins to circle around the origin
with an average radius $M(L)$. This motion is probably due to a random 
nonzero angular momentum of the initial configuration. Subtracting the 
angular momentum of OP, the relevant diffusion-like motion remains. 
MC studies employing first order Monte Carlo "time"-evolution of 
non-equilibrium one dimensional systems have shown that the sign for 
SSB is the exponential decrease of the diffusion constant with $L$
\cite{mukamel}. In accordance with the expectations based on the 
MWH-theorem we find in the present case that the diffusion constant
decreases according to a power law with the exponent $-1.16\pm0.1$.
}

{In conclusion we can state that the numerical study of the 
Hamiltonian dynamics
of the classical  O(2) symmetric scalar model in 2+1 dimensions
provides a non-trivial check of our understanding of real time relaxation
phenomena.}
Numerical results both for the late-time OP-asymptotics and for the
decay rate of on-shell waves (with well-defined wave vector $\bf k$)
were found to be in agreement with the
linear response theory. In the $h\neq 0$ equilibrium the
agreement of the masses coming from the 
perturbatively calculated effective potential with the numerically 
established excitation masses was also verified.
Finally, we have demonstrated the real time manifestation of 
the finite volume MWH-theorem by measuring the large $L$ asymptotics
of the angular diffusion rate of the macroscopic order parameter.

\subsection*{Acknowledgements}

We thank our supervisor A. Patk{\'o}s for his advice and constant
support and to A. Jakov\'ac for valuable and enlightening discussions.
We would like to thank to Z. R\'acz for his useful remarks on the issue
of MWH theorem. The authors also acknowledge the use of
computing resources provided by the Inst. for Theoretical Physics of
the E{\"o}tv{\"o}s University. 
This research received support of the Hungarian Science Fund (OTKA).

\end{document}